\documentclass[preprint,11pt]{elsarticle}
\usepackage[suffix=]{epstopdf}

\usepackage{geometry}
\usepackage{color}

\usepackage{amssymb,amsmath}
\usepackage{amsfonts}
\usepackage{graphics}

\newtheorem{theorem}{Theorem}
\newtheorem{lemma}[theorem]{Lemma}

\begin{document}

\title{Parameter Switching Synchronization}
\vspace{5mm}

\author [rm1,rm2]{Marius-F. Danca}
\author [rm3,rm4]{Nikolay Kuznetsov\corref{cor1}}
\cortext[cor1]{Corresponding author}
\address[rm1]{Dept. of Mathematics and Computer Science, Avram Iancu University of Cluj-Napoca, Romania}
\address[rm2]{Romanian Institute of Science and Tecçhnology, Cluj-Napoca, Romania}
\address[rm3]{Department of Applied Cybernetics, Saint-Petersburg State University, Russia}
\address[rm4]{Department of Mathematical Information Technology,
 University of Jyv\"{a}skyl\"{a}, Finland}

\begin{abstract}In this paper we show how the Parameter Switching algorithm, utilized initially to approximate attractors of a general class of nonlinear dynamical systems, can be utilized also as a synchronization-induced method. Two illustrative examples are considered: the Lorenz system and the Rabinovich-Fabrikant system.
\end{abstract}


\maketitle

\section{Introduction}

A number of various synchronization methods have been developed, such as complete or identical synchronization, phase and lag synchronization, generalized synchronization, intermittent lag synchronization, imperfect phase synchronization, almost synchronization and so on (see e.g. \cite{boca,peco,ivan}, or \cite{peco2}).

In this paper a new synchronization-induced method, which is based on the Parameter Switching (PS) algorithm \cite{danca}, is proposed. It is demonstrated \cite{dan,dan2} that this method can be effectively used for the approximation of attractors of a given nonlinear dynamical system which depends linearly on a real parameter.

Let us consider the following initial value problem (IVP), which models a large class of continuous-time nonlinear autonomous dynamical systems depending on a single real control parameter $p$, such as the Lorenz system, R\"{o}sler system, Chen system, Lotka--Volterra system, Rabinovich--Fabrikant system, Hindmarsh-Rose system, L\"{u} system, classes of minimal networks and many others, in the following form
\begin{equation}\label{e0}
\dot x(t)=f(x(t))+pAx(t),\quad x(0)=x_0,
\end{equation}
\noindent where $t\in I=[0,T]$, $x_0\in \mathbb{R}^n$, $p\in \mathbb{R}$ the control parameter, $A\in \mathbb{R}^{n\times n}$ a constant matrix, and $f : \mathbb{R}^n \rightarrow\mathbb{R}^n$ a nonlinear function.

For example, we can consider the IVP with $n=3$ for the Lorenz system
\begin{equation}%
\begin{array}
[c]{cl}%
\overset{\cdot}{x}_{1}= & \sigma(x_{2}-x_{1}),\\
\overset{\cdot}{x}_{2}= & x_{1}(\rho-x_{3})-x_{2},\\
\overset{\cdot}{x}_{3}= & x_{1}x_{2}-\beta x_{3},
\end{array}
\label{lorenz}%
\end{equation}
\noindent with $n=3$, $a=10$ and $c=8/3$, $p=\rho$ is the control parameter and
\[
f(x)=\left(
\begin{array}
[c]{c}%
\sigma(x_{2}-x_{1})\\
-x_{1}x_{3}-x_{2}\\
x_{1}x_{2}-\beta x_{3}%
\end{array}
\right)  ,~~A=\left(
\begin{array}
[c]{ccc}%
0 & 0 & 0\\
1 & 0 & 0\\
0 & 0 & 0
\end{array}
\right).
\]

The PS algorithm approximates numerically any solution of the IVP (\ref{e0}) \cite {danca,dan,dan2}. If one chooses a finite set of $N>1$ parameter values: $\mathcal{P}_N=\{p_1,p_2,...,p_N\}$, and one switch the parameter $p$ within $\mathcal{P}_N$ for a relatively short periods of time, while the underlying IVP is numerically integrated, then the resultant ``switched'' numerical solution will converge to the ``averaged'' solution, obtained for $p$ being replaced with the average of the switched values, given by
\begin{equation}\label{p}
p^*:=\frac{\sum_{i=1}^Nm_ip_i}{\sum_{i=1}^Nm_i},
\end{equation}

\noindent where $m_i$, $i=1,2,...,N$, are some positive integers, called ``weights''.\footnote{For some given $p^*$, the relation (\ref{p}) is verified for several other choices of $m_i$ $i=1,2,...,N$, and $\mathcal{P}_N$.}

We present in this paper how this algorithm can be used to obtain a synchronization between two systems modeled by the IVP (\ref{e0}), based on the convergence of the PS algorithm.

The paper is organized as follows: Section 2 describes the PS algorithm and how it can be implemented numerically, Section 3 presents the synchronization-induced by the PS algorithm and its application to the Lorenz and Rabinovich-Fabrikant systems. Conclusion  is summarized in the last section of the paper.

\section{Parameter Switching algorithm}
If, while the IVP (\ref{e0}) is integrated, $p$ is switched within $\mathcal{P}_N$, the obtained ``switching'' equation has the following form
\begin{equation}\label{e1}
\dot x(t)=f(x(t))+p(t)Ax(t),\quad x(0)=x_0,
\end{equation}

\noindent where $p:I\rightarrow\mathcal{P}_N$ is a piece-wise constant function that switches periodically its values
$p(t)=p_i$, $p_i\in\mathcal{P}_N$, $i\in\{1,2,...,N\}$ and the ``averaged'' equation of (\ref{e0}) (obtained for $p$ replaced with $p^*$ given by (\ref{p})), is
\begin{equation}\label{e2}
\dot {\bar x}(t)=f(\bar x(t))+p^*A\bar x(t),\quad t\in I=[0,T],\quad \bar x(0)=\bar x_0.
\end{equation}

\noindent Let us consider the following assumptions:

\noindent\textbf{Assumption H1}. The IVP (\ref{e0}) enjoys the uniqueness (e.g. $f$ satisfies the usual Lipschitz condition).

\noindent\textbf{Assumption H2}. The initial conditions $x_0$ and $\overline{x}_0$ of (\ref{e1}) and (\ref{e2}), respectively, belong to the same basin of attraction $\mathcal{V}$ of the solution of (\ref{e2}).

\noindent Then, the relation between the solutions of (\ref{e1}) and (\ref{e2}) is given by the following lemma \cite{dan,dan2}

\begin{lemma}\label{lem}
For any close initial conditions $x_0, \overline{x}_0\in \mathcal{V}$, the ``switched'' solution approximates the ``averaged'' solution.
\end{lemma}

\noindent In \cite{dan} the proof of Lemma \ref{lem} is made on the basis of the global error of Runge-Kutta, while in \cite{dan2} the average theory \cite{ver} has been utilized. The proof can be also done constructively, using nonlinear tools like Poincar\'{e} sections, time series analysis, histograms \cite{danca}.

As common in numerical approach of nonlinear systems, throughout this paper, for some given $p$ and $x_0$, the ``attractors'' are the considered as the numerical approximation of $\omega$-limit sets \cite{foi}, neglecting sufficiently long transients. Then, thanks to the convergence of the PS algorithm (Lemma \ref{lem}), every attractor of the underlying system can be numerically approximated by the algorithm.

As is well known, attractors present continuous dependence on the underlying parameter. This is roughly speaking, the dependence of the solution of the IVP (\ref{e0}) on $p$
is continuous as long as the function $f$ is continuous (see \cite{hum} or \cite{perk}, p. 83).

To implement numerically the PS algorithm, a fixed single step numerical method (such as the standard Runge-Kutta (RK4) method utilized here, with the step size $h$) is necessary. Symbolically, the algorithm is described by the following scheme:
\begin{equation}\label{scheme}
[m_1p_1,m_2p_2,...,m_Np_N],
\end{equation}

\noindent which means that while the IVP is integrated, for $m_1$ integration steps $p$ is set to $p_1$, then for the next $m_2$ steps, $p=p_2$, and so on, until the last $m_N$ steps, with $p=p_N$. Next, the algorithm repeats periodically (with period $(m_1+m_2+...+m_N)h$), for the next set of $N$ values of $p$, following the same rule, until the integration time interval $I$ is completely covered.

The strongness of the PS algorithm lies, mainly, on the linear dependence on $p$ of the right-hand side of the system (\ref{e0}) and on the convexity of the relation (\ref{p}): denoting $\alpha_j=m_j/\sum_{i=1}^N m_i$, $j=1,2,...,N$, the relation (\ref{p}) becomes $p^*=\sum_{i=1}^N\alpha_i p_i$ with $\sum_{i=1}^N\alpha_i=1$. Therefore, for any set $\mathcal{P}_N$ and weights $m_i$, $i=1,2,...,N$, $p^*$ belongs always inside the interval $(p_{min},p_{max})$, where $p_{min} \equiv min\{\mathcal{P}_N\}$ and $p_{max} \equiv max\{\mathcal{P}_N\}$. Therefore, in numerical experiments, to approximate some attractor $A_{p^*}$ with the PS algorithm, the set $\mathcal{P}_N$ has to be chosen such as $p^*\in(p_{min},p_{max})$.

\section{Synchronization-induced by the PS algorithm}

\begin{figure}
\begin{center}
   \includegraphics[clip,width=0.5\textwidth]{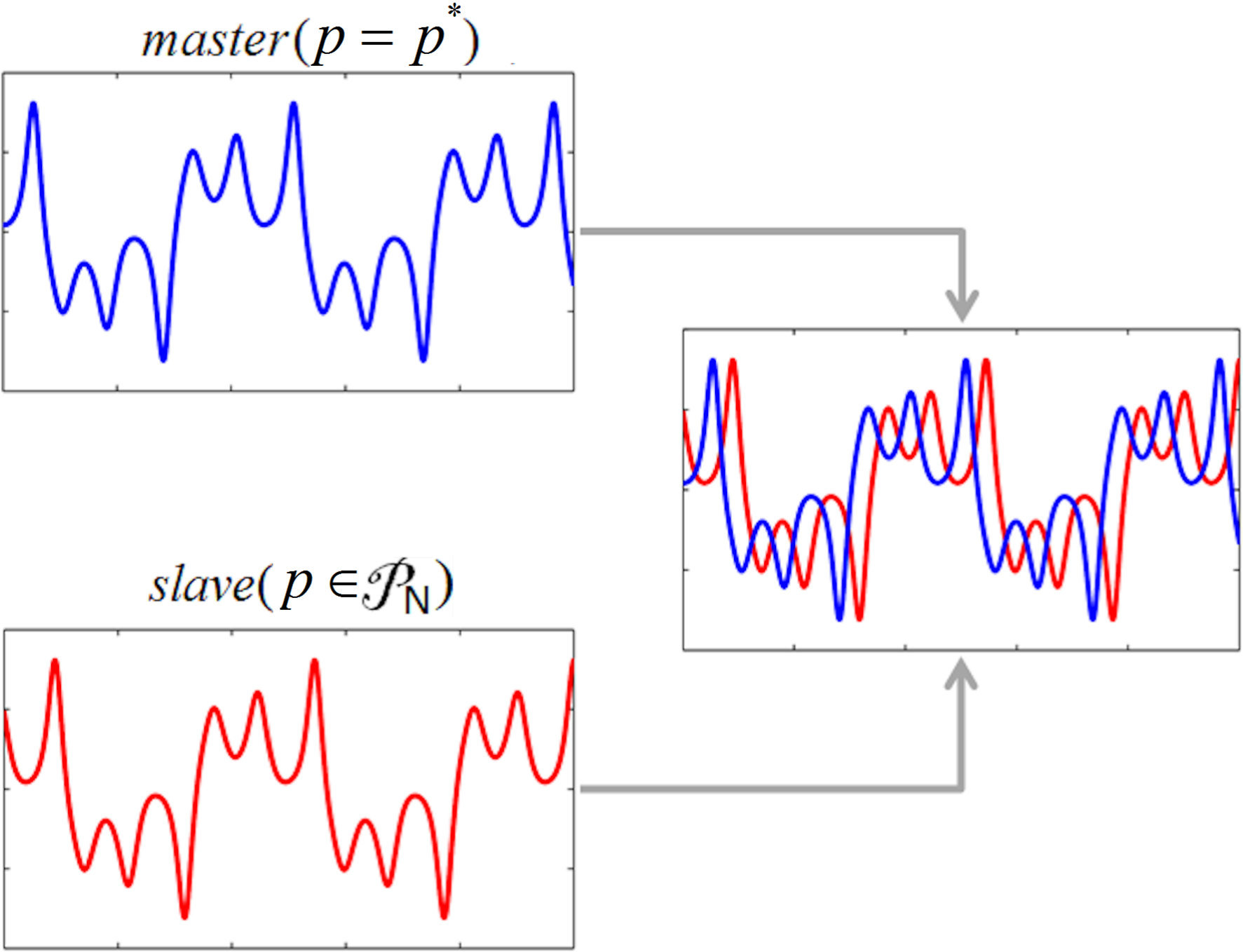}
\caption{PSLS algorithm, a sketch.}
\label{fig1}
\end{center}
\end{figure}

Based on Lemma \ref{lem}, consider a master-slave synchronization of two identical systems. Suppose that the master system evolves along a stable attractor, an oscillatory orbit ($A_{p^*}$) corresponding to $p = p^*$. Obviously, if the slave system has the knowledge of this particular parameter value and can use it, then the slave system, for the same initial conditions, will become just a copy of the master system thereby they will behave the same. In practice, however, this is unlikely the case but more generally the slave system has to estimate or to ``learn'' about this particular parameter value, thereby generating the stable cycle ($A^*$) to follow as close as possible (i.e. synchronize with) the master system which evolves along the attractor $A_{p^*}$. Here, via Lemma \ref{lem}, this task can be implemented with the PS algorithm (Fig. \ref{fig1}). For this purpose, we have to choose for the slave system the set $\mathcal{P}_N$ with the weights $m_i$, $i=1,2,...,N$, such that (\ref{p}) is verified. By using the PS algorithm, the stable (``switched'') attractor $A^*$ is generated, which is a very good approximation of the master cycle (``averaged attractor'') $A_{p^*}$. As known, the lag synchronization of chaotic systems implies that the state variables of the two coupled systems are synchronized but with a time lag with respect to each other (see e.g. \cite{lag,lag2,lag3,lag4}). As the sketch in Fig. \ref{fig1} indicates, the PS algorithm, induces a time lag between the synchronized stable orbits. Therefore, hereafter this synchronization method is called Parameter Switching Lag Synchronization (PSLS).

\vspace{3mm}

\noindent\textbf{Application}

The PSLS numerical tests in this paper have been realized via the standard RK4 with the integration step size $h=0.001$ and for the difference between the initial conditions ($x_0$ and $\bar x_0$) of order  $10^{-2}$ (Assumption \textbf{H2}). To emphasize the perfect PSLS, phase overplots, Poincar\'{e} overplotted sections, overplotted time series and simultaneous plots for the master (blue) and slave (red) synchronized attractors have been determined. For the considered examples, the Hausdorff distance (\cite{falco}, p.114) between the two synchronized (switched and averaged) attractors after transients are removed was $\approxeq 10^{-5}$. Limitations of the PSLS algorithm, are mainly due to the utilized numerical method to integrating the IVP, finite precision of computations involving floating-point, or $p^*$ infinite (repeating) decimals \cite{dan}.

\vspace{3mm}
\noindent\emph{Lorenz system}

\begin{figure}
\begin{center}
  \includegraphics[clip,width=0.9\textwidth] {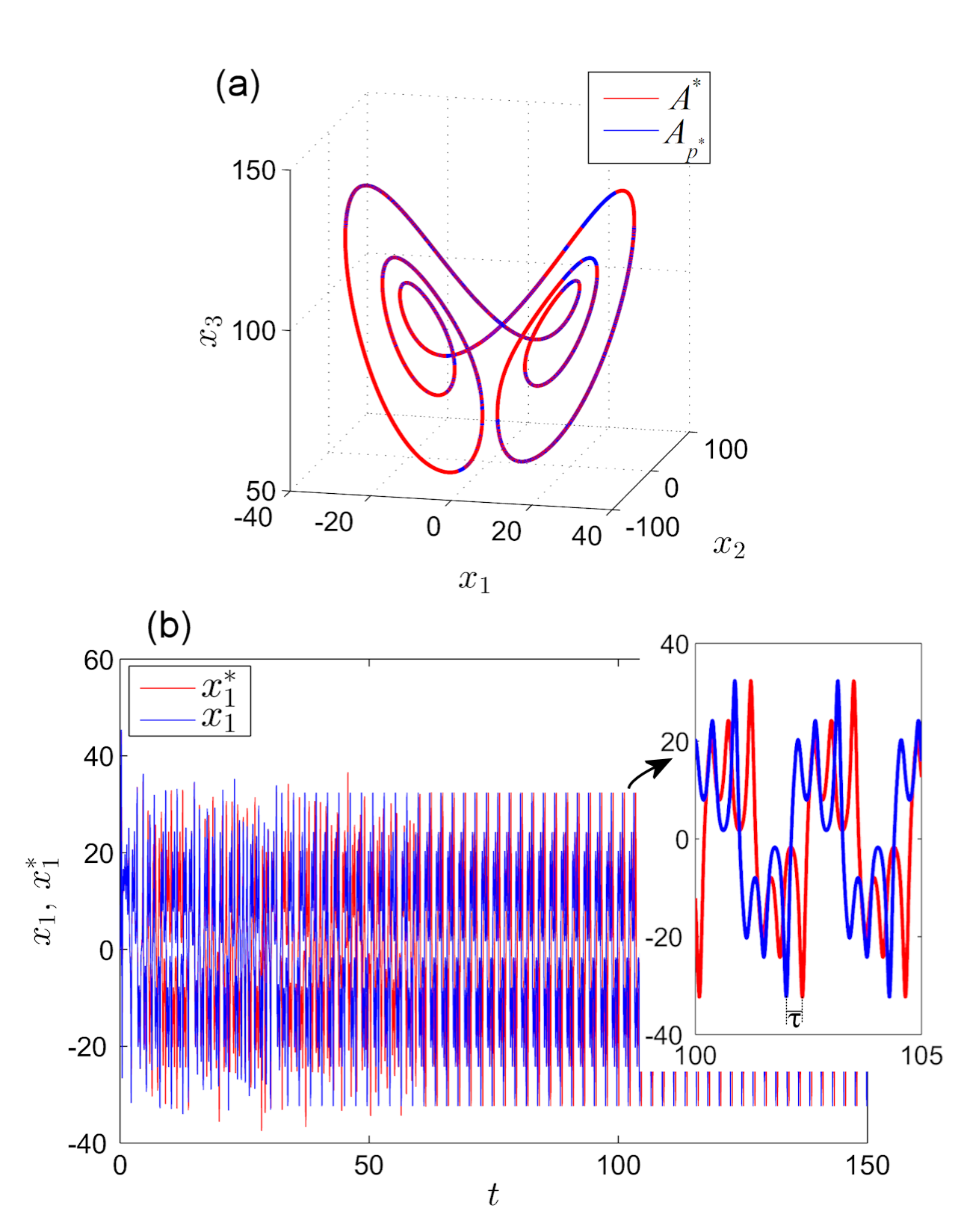}
\caption{PSLS for Lorenz system, for $p^*=93$, using the scheme $[1p_1,1p_2]$ with $p_1=90$ and $p_2=96$. (a) Phase overplots of the synchronized cycles. (b) Time series overplots of the first components $x_1$ and $x_1^*$ revealing the lag $\tau$ between the two cycles.}
\label{fig2}
\end{center}
\end{figure}

\begin{figure}
\begin{center}
  \includegraphics[clip,width=0.9\textwidth]{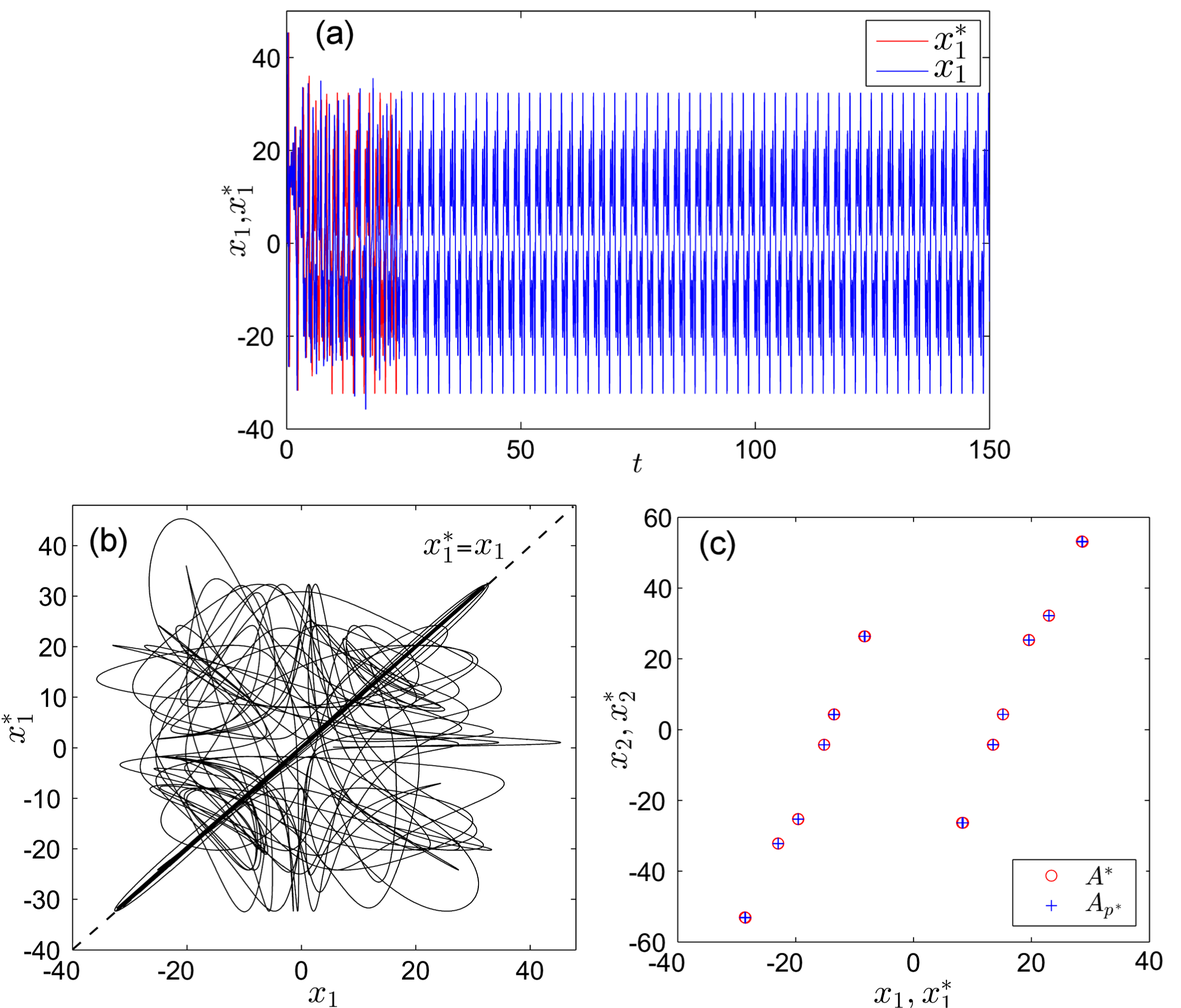}
\caption{PSLS for Lorenz system, for $p^*=93$, after lag removing. (a) Time series overplots of the first components $x_1$ and $x_1^*$. (b) Simultaneous plots of the first components $x_1$ and $x_1^*$ after lag removing. (c) Overploted Poincar\'{e} sections on the plane $x_3=100$. }
\label{fig3}
\end{center}
\end{figure}

Suppose that the Lorenz (master) system evolves along the stable cycle corresponding to $p=p^*=93$ (Fig. \ref{fig2} (a)) and we want to synchronize the slave system with this cycle. Let $N=2$ parameter values $\mathcal{P}_2=\{90,96\}$ and the wights $m_1=m_2=1$. Then, the relation (\ref{p}) gives $p^*=\frac{1 \times 90+1 \times 96}{1+1}=93$. Therefore, after some transients (Fig. \ref{fig2} (b)), the PS algorithm, applied periodically via the scheme (\ref{scheme}) (with period $(m_1+m_2)h=2h$), yields the PSLS. While Fig. \ref{fig2} (a) reveals a perfect match between the two stable periodic motions in the phase space, namely the averaged attractor $A_{93}$ (blue) and the switched following cycle $A^*$ (red), the time series of the first components $x_1$ and $x_1^*$ in Fig. \ref{fig2} (b) indicates the lag $\tau$ existence between the two cycles. After removing the lag, the two systems present a perfect PSLS (ensured by the algorithm convergence), as indicated by the time series (Fig. \ref{fig3} (a)), the simultaneous plot (Fig. \ref{fig3} (b)) and the overplotted Poincar\'{e} sections on the plane $x_3=100$ (Fig. \ref{fig3} (c)) (for the sake of clarity, on the Poincar\'{e} sections the transients have been removed). The same synchronization can be realized with several schemes (\ref{scheme}). For example, with $P_4=\{85,92,94,96.5\}$ and weights $m_1=2$, $m_2=1$, $m_3=3$ and $m_4=4$, the relation (\ref{p}) gives the same value $p^*=93$.

\begin{figure}
\begin{center}
  \includegraphics[clip,width=0.9\textwidth]{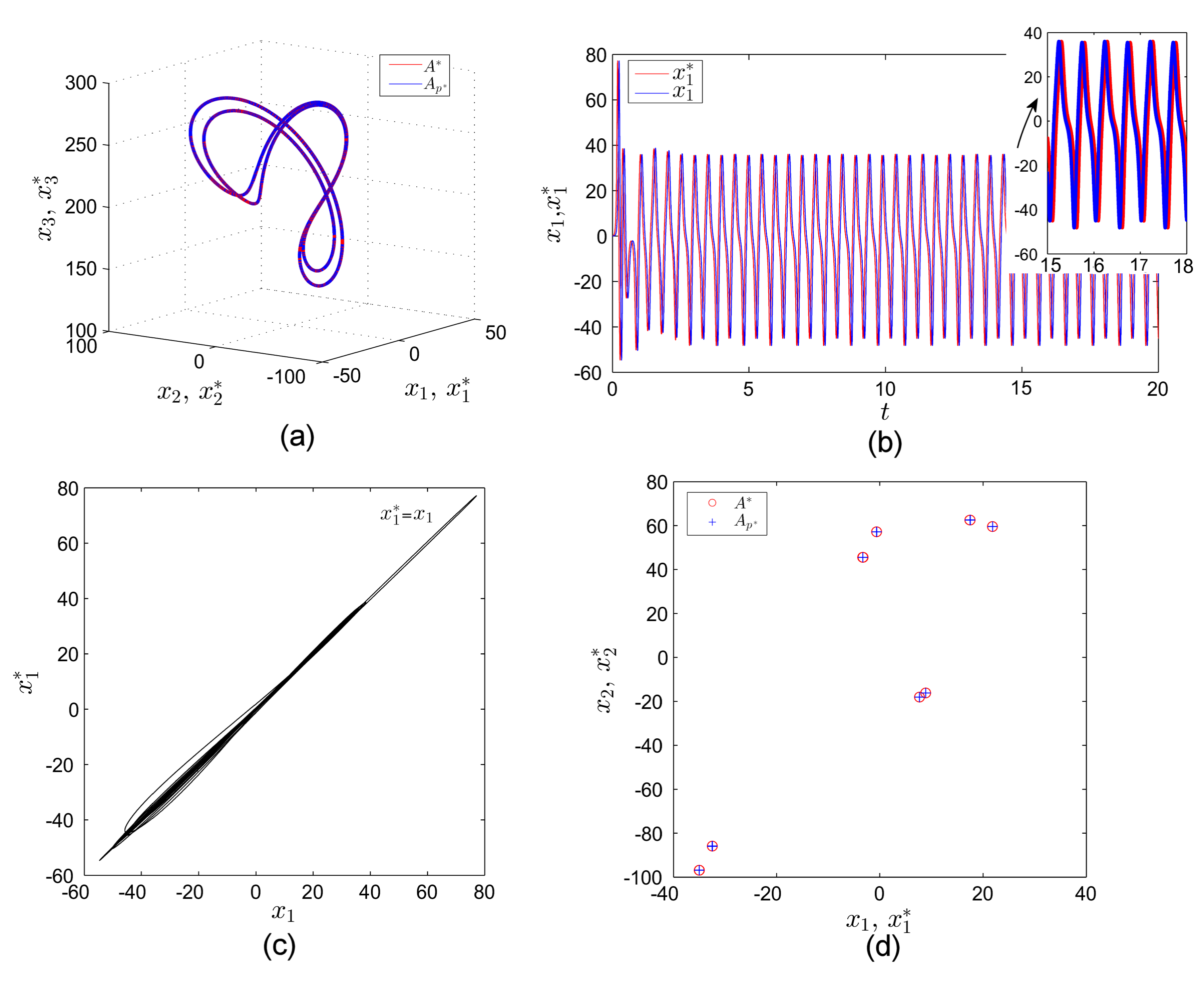}
\caption{PSLS of two stable Lorenz cycles for $p^*=220$. (a) Phase overplots of the synchronized cycles. (b) Time series overplots of the first components $x_1$ and $x_1^*$, with zoomed region revealing a small time lag between time series. (c) Simultaneous plots of the first components $x_1$ and $x_1^*$ after lag removing. (c) Overploted Poincar\'{e} sections on the plane $x_3=210$. }
\label{fig4}
\end{center}
\end{figure}

Consider next the case when one wants the slave system follow the stable cycle $A_{220}$ of the master system (see Fig. \ref{fig4} (a)). Then one can use for example, the scheme $[1p_1,2p_2,1p_3,2p_4,1p_5]$, with $\mathcal{P}_5=\{200,205,210,236,248\}$, which gives $p^*=220$. In this case, as can be seen in Fig. \ref{fig4} (b), due to a stronger attraction force of $A_{220}$ (compared to the case of $A_{93}$), the synchronization time is shorter (see the shorter transient in Fig. \ref{fig4} (b) and simultaneous plots in Fig. \ref{fig4} (c) which close to the line $x_1=x_1^*$ after only few integration steps). Also the lag between the two time series is smaller (detail in Fig. \ref{fig4} (b)). After removing the transients, the overploted Poincar\'{e} sections with the plane $x_3=320$ reveal the perfect match between $A_{220}$ and $A^*$.

\vspace{3mm}
\noindent\emph{Rabinovich-Fabrikant system}

Let us next consider a system with strong nonlinearities, the Rabinovich-Fabrikant (RF) system, modeled by the following IVP \cite{danx}

\begin{equation}
\label{rf}
\begin{array}{l}
\overset{.}{x}_{1}=x_{2}\left( x_{3}-1+x_{1}^{2}\right) +ax_{1}, \\
\overset{.}{x}_{2}=x_{1}\left( 3x_{3}+1-x_{1}^{2}\right) +ax_{2}, \\
\overset{.}{x}_{3}=-2x_{3}\left(p+x_{1}x_{2}\right),
\end{array}%
\end{equation}

\begin{figure}
\begin{center}
  \includegraphics[clip,width=0.9\textwidth]{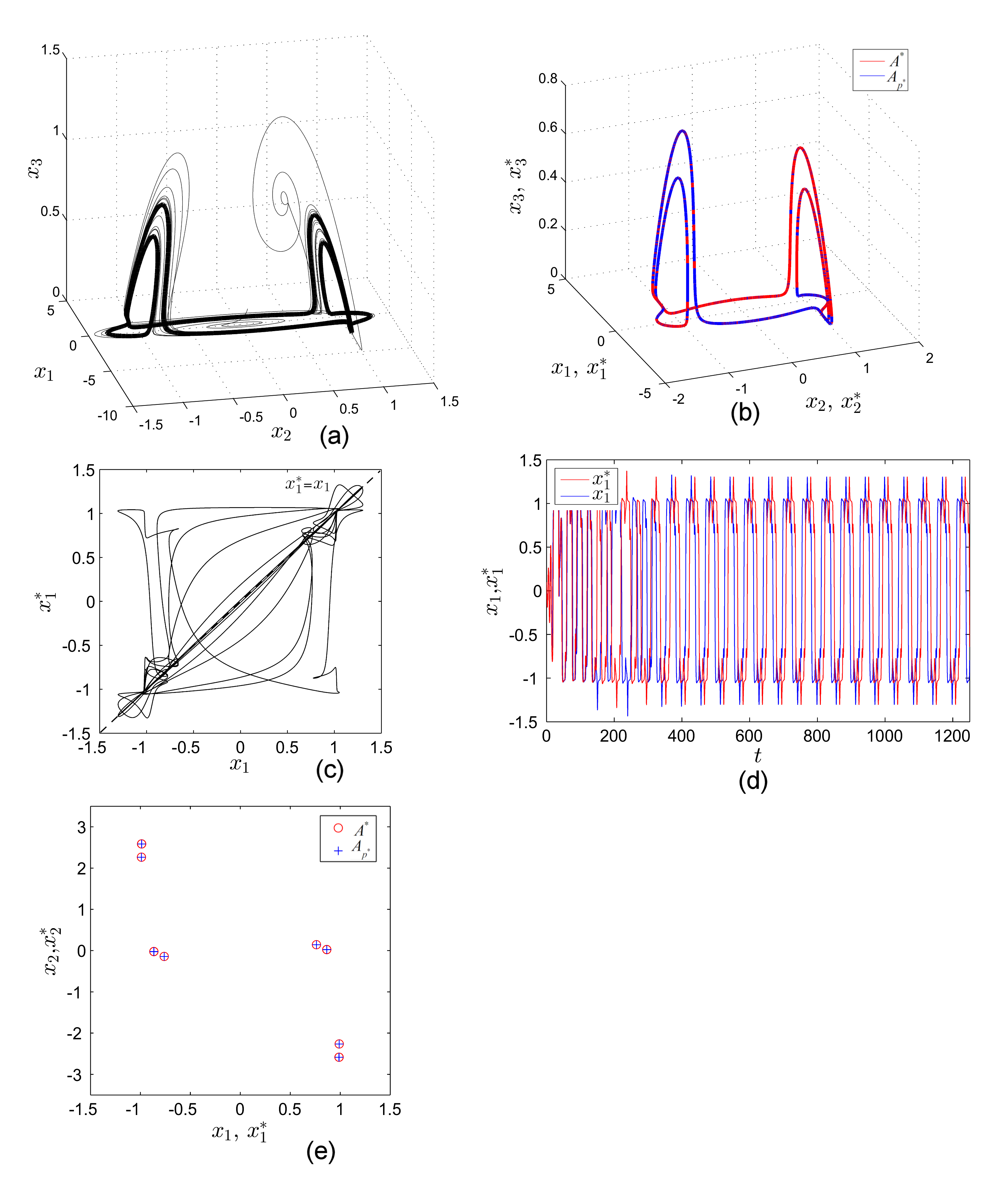}
\caption{PSLS of two stable cycles of the RF system (\ref{rf}) corresponding to $a=0.1$ and $p^*=1.035$. (a) The stable cycle for $p^*=1.035$ (black). Transients (grey) indicates the existence of a heteroclinic orbit. (b) Phase overplots of the synchronized cycles. (c) Simultaneous plots of the first components $x_1$ and $x_1^*$ after lag removing. (d) Time series overplots of the first components $x_1$ and $x_1^*$. (e) Overploted Poincar\'{e} sections on the plane $x_3=0.35$. }\label{fig5}
\end{center}
\end{figure}

\noindent Suppose one intends to synchronize a slave RF system (\ref{rf}), for $a=0.1$, with the master system which evolves on the stable cycle $A_{1.035}$ (see Fig. \ref{fig5} (a), where the transients (grey) reveal the rich system dynamics, such as hyperbolic orbits before reaching $A_{1.035}$ (see \cite{danx} for several properties of this system)). For this purpose one can use, for example, the scheme $[2p_1,1p_2,3p_3]$ with $p_1=1$, $p_2=1.03$, $p_3=1.06$ (Fig. \ref{fig5}). Due to the slow attractor speed, the time series reveals the necessary longer integration time Fig. \ref{fig5} (d). The phase overplots, simultaneous plots and overplotted Poincar\'{e} sections with the plane $x_3=0.1$ (Fig. \ref{fig5} (b), (c) and (e) respectively), reveal the perfect PSLS.

\begin{figure}
\begin{center}
  \includegraphics[clip,width=0.9\textwidth] {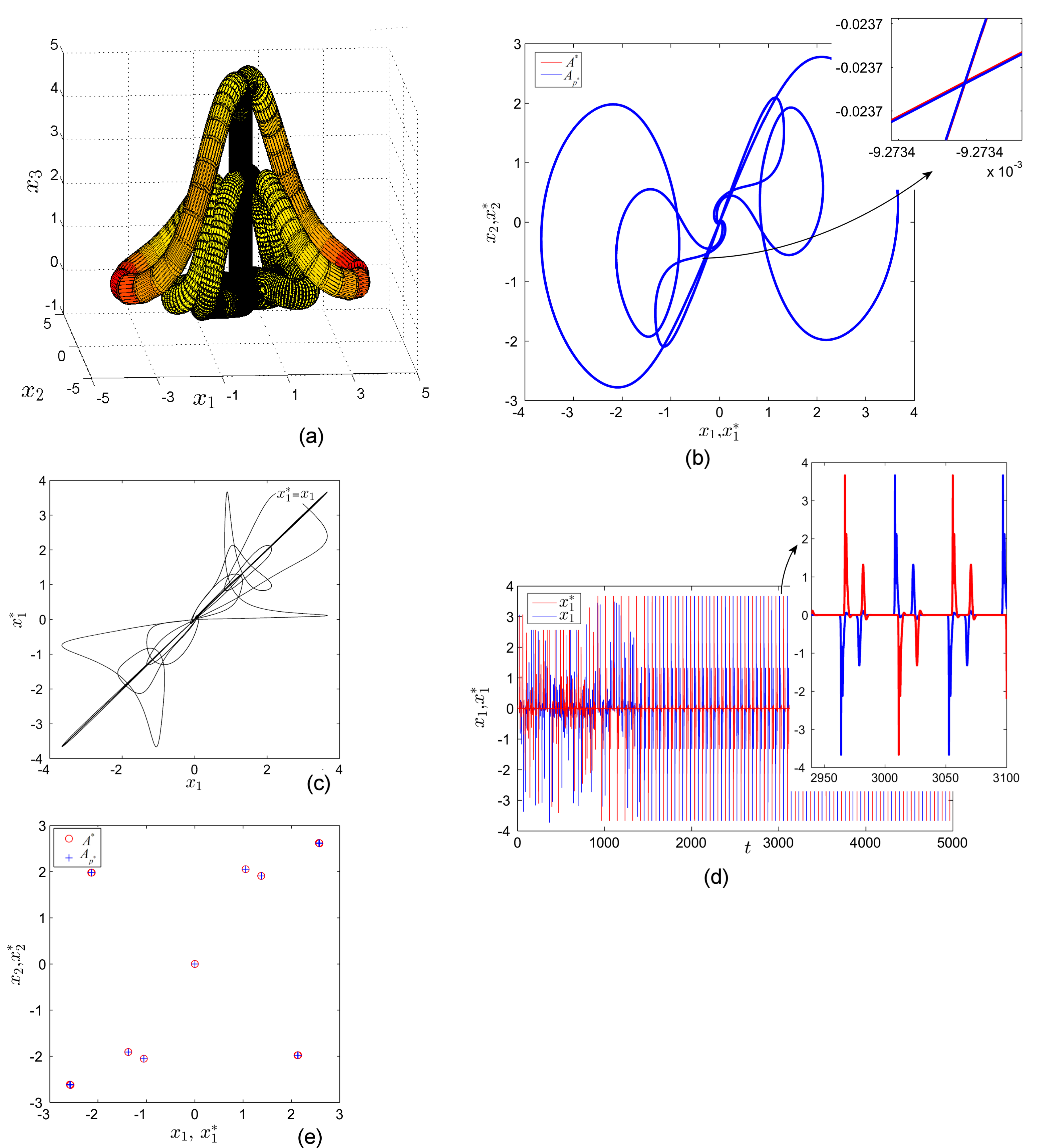}
\caption{PSLS of two stable cycles of the RF system (\ref{rf}) corresponding to $a=-1$ and $p^*=-0.1$. (a) Tubular phase representation. (b) Plane phase overplots of the synchronized cycles. (c) Simultaneous plots of the first components $x_1$ and $x_1^*$ after lag removing. (d) Time series overplots of the first components $x_1$ and $x_1^*$. Zoomed region indicates the relatively large time lag. (e) Overploted Poincar\'{e} sections on the plane $x_3=1$. }
\label{fig6}
\end{center}
\end{figure}

Another stable cycle of the RF system with complicated dynamics, corresponds to $a=-1$ and $p^*=-0.1$, depending strongly on the numerical method, initial conditions and the step size \cite{danx} (see the tubular representation in Fig. \ref{fig6} (a); red color represents the higher system speed along the cycle). By using the scheme $[1p_1,1p_2]$ with $p_1=-0.15$ and $p_2=-0.05$, the results of the PSLS is presented in Fig. \ref{fig6} (b)-(e).
The zoomed region drawn in the phase plane $(x_1,x_2)$ (Fig. \ref{fig6}) (b)) reveals the perfect match between the two cycles. To note the relatively large lag in this case (Fig. \ref{fig6} (d)). Simultaneous plots of the first components (Fig. \ref{fig6} (c) and overploted Poincar\'{e} sections on the plane $x_3=1$ (Fig. \ref{fig6} (e)) underlines the rightness of the PSLS.

Even this method acts differently to the classical synchronization methods, it presents several advantages such as: it does not require calculating the Jacobian, the coupling parameter, and integration of both master and slave systems. The only requirement is the knowledge of the targeted parameter value $p^*$ and a set of accessible values $\mathcal{P}_N$ such as $p^*\in(p_{min},p_{max})$.

\begin{figure}
\begin{center}
  \includegraphics[clip,width=0.9\textwidth] {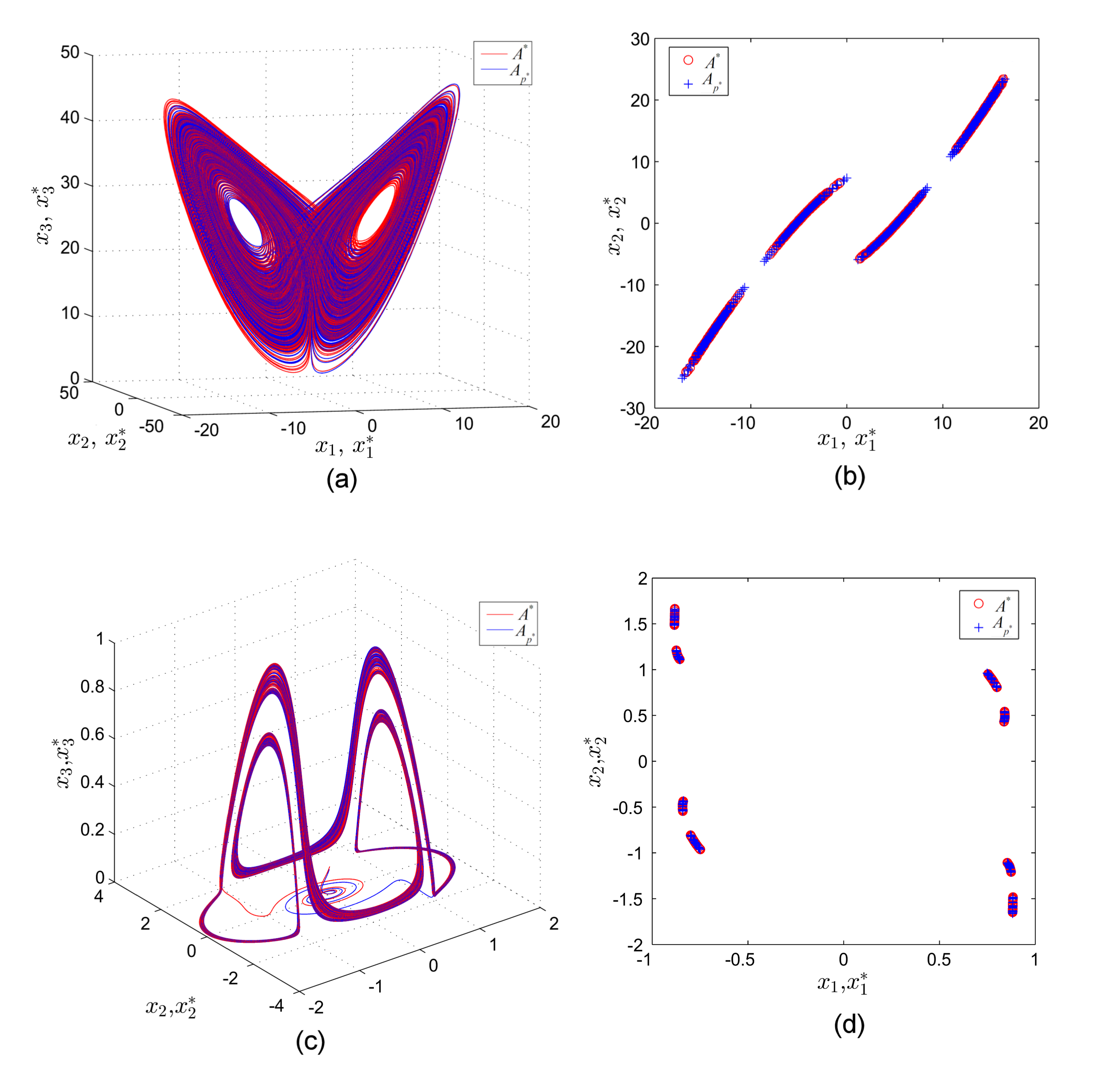}
\caption{(a) PSLS of two chaotic Lorenz attractors, for $p^*=28$ (phase overplots). (b) Overploted Poincar\'{e} sections on the plane $x_3=30$. (c) PSLS of two chaotic RF attractors, corresponding to $p^*=0.2876$ (phase overplots). (b) Overploted Poincar\'{e} sections on the plane $x_3=0.35$.}
\label{fig7}
\end{center}
\end{figure}

Our numerous numerical tests reveal the fact that chaotic motions could also be synchronized with the PSLS algorithm in the following sense. Let us consider for example a chaotic attractor $A_{p^*}$ of the Lorenz system with, e.g., $p^*=28$. The scheme $[1p_1,1p_2]$ with $p_1=26$ and $p_2=30$, gives the desired value $p^*=28$. After applying the PSLS, the results are presented in Fig. \ref{fig7}. The phase overplots (Fig. \ref{fig7} (a)) and overplotted Poincar\'{e} sections on the plane $x_3=30$ (Fig. \ref{fig7} (b)), reveal that the two attractors tend to cover finally the same path in the phase space. For the chaotic RF attractor, corresponding to $p^*=0.2876$, with the scheme $[1p_1,2p_2,2p_3]$ with $p_1=0.28$,$p_2=0.288$, $p_3=0.291$, one obtain the PSLS in Fig. \ref{fig7} (c), (d). As one can see, the phase plots (Fig. \ref{fig7} (a) and (c)) reveal some apparently difference between the synchronized attractors. However, as known, chaotic attractors, require theoretically an asymptotically (infinitely) long time compared with the finite time used in numerical simulations. Another reason for this apparent difference is the lag (which is more difficult to determine in the chaotic synchronization).

Not only periodic ways to implement the scheme (\ref{scheme}) can be implemented, but also some random ways can be used to realize synchronization \cite{foot2}.

\section{Conclusion}
In this paper we presented another synchronization method based on the PS algorithm when a suitable set of switching parameter values is chosen. The good convergence of the numerical solution of the slave system, subjected to the switching of $p$, to the corresponding solution with the average value of $p$ (i.e. $p^*$ in the master system), and the ease to implement the PSLS (e.g. no Jacobian matrix is involved), make the PSLS a promising new synchronization-induced method.
Deepen underlying lagged relationships, especially for chaotic attractors synchronization with the PSLS method represents an important future task. Thus, a promising approach to improve and clarify the chaotic PSLS, would be the ``system'' approach, in which one of the time series is viewed as input and the other series is considered as output.

\bigskip
\noindent
{\bf Acknowledgment} NK is supported by the Saint-Petersburg State University.

\newpage

\end{document}